\documentclass[review]{elsarticle}
\usepackage{lineno,hyperref}
\usepackage{amsmath}
\usepackage{amssymb}
\journal{Journal of \LaTeX\ Templates}

\begin{document}

\begin{frontmatter}

\title{A prototype detector for the CRESST-III low-mass dark matter search}

\author[a]{R. Strauss\corref{mycorrespondingauthor}}

\cortext[mycorrespondingauthor]{Corresponding author}
\ead{strauss@mpp.mpg.de}
\author[a]{G. Angloher}

\author[a]{P. Bauer}




\author[b]{X. Defay}

\author[b,e]{A. Erb}

\author[b]{F.v. Feilitzsch}

\author[a]{N. Ferreiro\,\,Iachellini}



\author[b]{R. Hampf}

\author[a]{D. Hauff}


\author[a]{M. Kiefer}



\author[b]{J.-C. Lanfranchi}
\author[b]{A. Langenk\"amper}

\author[b]{E. Mondragon}

\author[b]{A. M\"unster}

\author[b]{C. Oppenheimer}

\author[a]{F. Petricca}

\author[b]{W. Potzel}

\author[a]{F. Pr\"obst}

\author[a]{F. Reindl}

\author[a]{J. Rothe}




\author[b]{S. Sch\"onert}

\author[a]{W. Seidel}
	
\author[b]{H. Steiger}
\author[a]{L. Stodolsky}


\author[a]{A. Tanzke}

\author[b]{H.H. Trinh Thi}

\author[b]{A. Ulrich}


\author[b]{S. Wawoczny}

\author[b]{M. Willers}

\author[a]{M. W\"ustrich}

\author[b]{A. Z\"oller}


\address[a]{Max-Planck-Institut f\"ur Physik,   D-80805 M\"unchen, Germany}
\address[b]{Physik-Department, Technische Universit\"at M\"unchen, D-85748 Garching, Germany}
\address[e]{Walther-Mei\ss ner-Institut f\"ur Tieftemperaturforschung,  D-85748 Garching, Germany}




\begin{abstract}
The CRESST-III experiment which is dedicated to low-mass dark matter search uses scintillating CaWO$_4$ crystals operated as cryogenic particle detectors.  Background discrimination is achieved by exploiting the scintillating light signal of CaWO$_4$ and by a novel active detector holder presented in this paper. In a test setup above ground, a nuclear-recoil energy threshold of $E_{th}=(190.6\pm5.2)\,$eV is reached with a 24\,g prototype detector, which  corresponds to an estimated threshold of $\sim$50\,eV when being operated in the low-noise CRESST cryostat.  This is the lowest threshold reported for direct dark matter searches.  For CRESST-III phase 1, ten such detector modules were installed in the cryostat  which  have the potential to improve significantly the sensitivity to scatterings of dark matter particles with masses down to $\sim$0.1\,GeV/c$^2$.
\end{abstract}


\end{frontmatter}


\section{Introduction}
Although plenty of evidence exists for dark matter (DM) in the Universe, direct searches so far did not observe an unambiguous signal \cite{Klasen20151}. In particular, for low-mass DM particles with masses of 1-10\,GeV large parts of the accessible parameter space of spin-independent cross section for scatterings on nuclei remain untested, despite many naturally motivated theoretical models for light dark matter, such as e.g. asymmetric DM \cite{asymmReview}. 

The CRESST experiment (Cryogenic Rare Event Search with Superconducting Thermometers) currently provides the best sensitivity for scatterings of DM particles with masses below $\sim$2\,GeV \cite{Angloher:2014myn,Angloher:2015ewa}.  The detectors operated in  CRESST-II phase 2 reached thresholds down to $\sim$300\,eV. These recent results showed that the nuclear-recoil energy threshold is the main driver for low-mass DM search. For particles of mass $m_\chi$ scattering on a nucleus of mass number $A$ a maximum energy transfer of 
\begin{equation}
E_{max}\approx 130 \left(\frac{m_\chi}{1\,\mathrm{GeV/c}^2}\right)^2\left(\frac{100}{A}\right)\,\mathrm{eV}
\end{equation}  
is expected under standard assumptions \cite{Klasen20151}.  Hence, for the next phase of the experiment (CRESST-III), which is dedicated to low-mass DM search, detectors with an extremely low energy threshold are necessary. For CRESST-III phase 1, a straight-forward approach was chosen: the CRESST-II detectors with a mass of $\sim$300\,g are scaled down in size to $\sim$24\,g. Thereby, an improvement of the signal-to-noise ratio of up to a factor $\sim$10 is expected due to basic phonon physics \cite{Probst:1995fk}. In this paper, the design and performance of a prototype detector is presented which shows that nuclear-recoil energy thresholds of $\sim$50\,eV are in reach.   


\section{Detector design}
\begin{figure}
\centering
\includegraphics[width=0.7\textwidth]{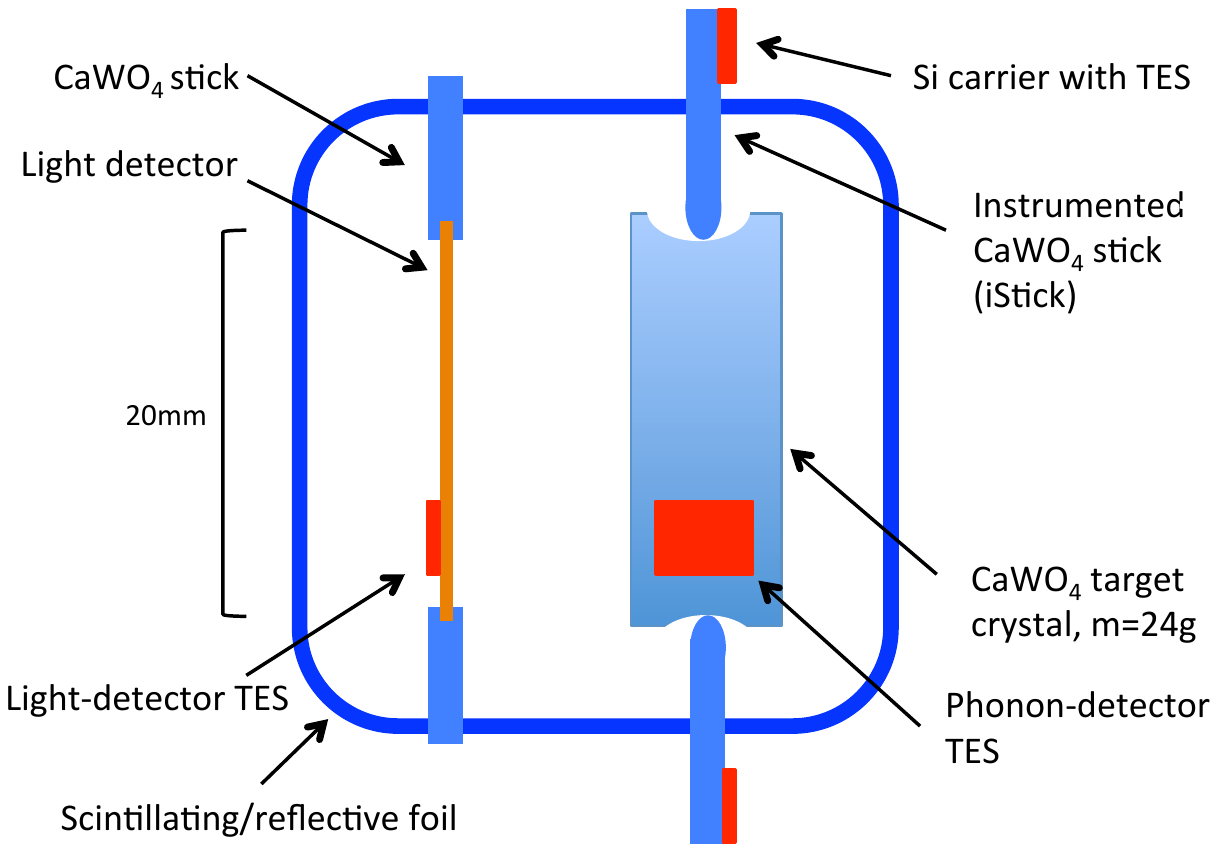}
\caption{Schematic view of the CRESST-III detector module.}
\label{fig:CRESST3_detector}
\end{figure}

The CRESST-III detector module consists of a 24\,g CaWO$_4$ crystal (20x20x10\,mm$^3$) as target material, called phonon detector (PD), and a silicon-on-sapphire disc (20x20x0.4\,mm$^3$) as a light absorber (see Fig. \ref{fig:CRESST3_detector}), called light detector (LD). This 2-channel readout provides an active background discrimination: while for electron/gamma events a fraction of $\sim$5\% of the deposited particle energy is converted into scintillation light, for nuclear recoils this fraction is reduced due to light quenching \cite{birks1964theory}. Depending on the recoiling nucleus 2-12\% \cite{Strauss:2014ab} light is emitted compared to electron/gamma events of the same energy.  Both detectors are equipped with transition-edge-sensors (TES) realized by thin W films which are operated in the transition between the superconducting and normal conducting state \cite{Probst:1995fk}. The resistance change caused by a phonon-induced temperature rise in the crystal is measured with a SQUID system \cite{Angloher:2012vn}.   While the design of the LD-TES is unchanged with respect to earlier CRESST measuring campaigns, the TES of the PD was significantly modified and optimized to meet the requirements of low-mass DM search. The detectors are held by CaWO$_4$ sticks (3 each), a design which was already successfully implemented in CRESST-II phase 2 \cite{Angloher:2014myn,strauss:2014part1}. In addition, a novel concept for an instrumented detector holder is introduced: the sticks holding the PD are equipped with TESs which realize a veto against events related to the crystal support.   

\subsection{CaWO$_4$ cryogenic detector}
\begin{figure}
\centering
\includegraphics[width=0.7\textwidth]{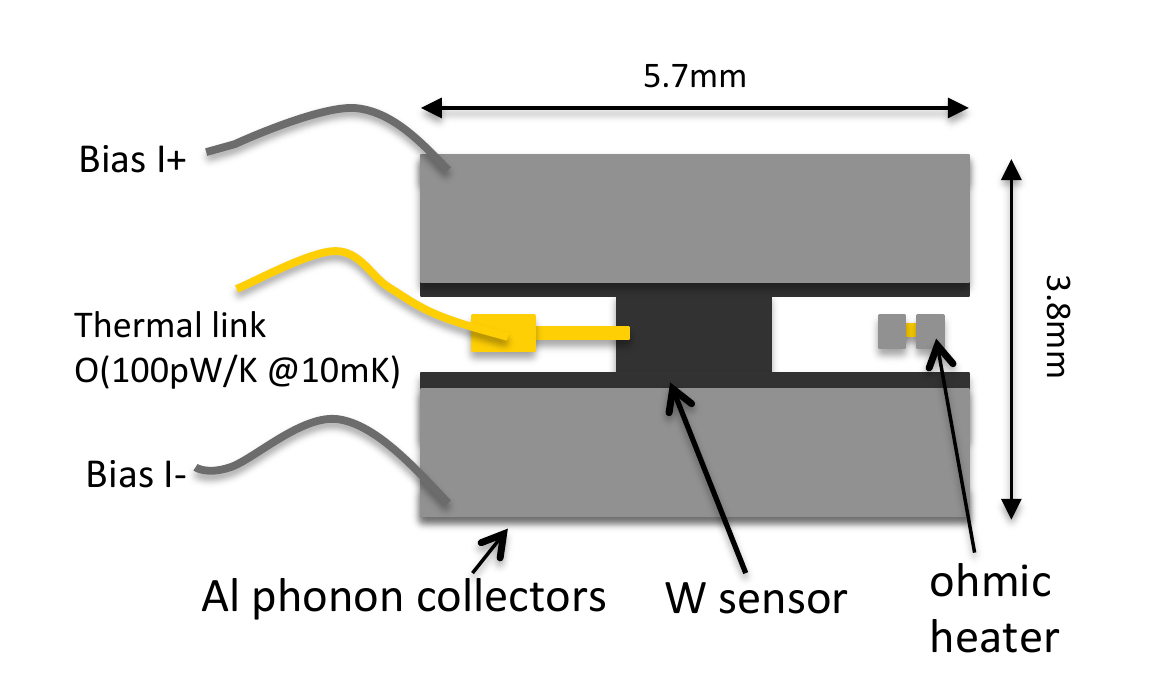}
\caption{Schematic view of the TES of the CaWO$_4$ PD for CRESST-III. The W-film is equipped with Al-phonon collectors and a Au-stripe as weak thermal link. A separated ohmic heater is used to apply artificial pulses for calibration. The electrical and thermal contacts to the holder are provided by Al and Au wire bonds (diameter 25\,$\mu$m). }
\label{fig:TES}
\end{figure}

The TES of the CaWO$_4$ detector is designed such that the detector is operated in the calorimetric mode \cite{Probst:1995fk}, i.e. the sensor integrates over the impinging non-thermal phonons created by a particle interaction in the crystal. This is achieved by a 2.4x0.85\,mm$^2$ W film (thickness 200\,nm) as thermometer which is weakly coupled to the heat bath (copper detector holder) by a thermal link ($\sim$100\,pW/K at 10\,mK). The thermal coupling is  realized by a Au stripe (1.0x0.02\,mm$^2$, thickness: 20\,nm) which is sputtered onto the crystal. A scheme of the TES is shown in  Fig. \ref{fig:TES}. The collection area of the sensor for phonons can be increased - without increasing the heat capacity of the thermometer - by superconducting Al phonon collectors. Incident phonons break cooper pairs in the Al which can diffuse into the W-film, couple to its electron system and thereby increase the temperature signal \cite{Angloher2016}. A separate ohmic heater (Au film)  is used to inject artificial heater pulses for calibration and stabilization \cite{Angloher:2012vn}.

The CaWO$_4$ crystal (\textit{TUM56f}) used for the prototype module  was grown in-house at the Technische Universit\"at M\"unchen (TUM) \cite{erb}. It is expected to have similar properties in terms of radiopurity and optical quality as the crystal TUM40 operated in CRESST-II phase 2 \cite{strauss:2014part1,strauss:2014part2} since it was produced under similar conditions. A mean background level of  3.51\,events/[kg\,keV\,day] at 1-40\,keV was observed with TUM40 which is about an order of magnitude better than previously used commercial CaWO$_4$ crystals.
The surfaces of the crystal are roughened in order to increase the scintillation light output. The dips where the sticks touch (see Fig. \ref{fig:CRESST3_detector}) are polished to optical quality to avoid thermal stress at the contact points. 

\subsection{Instrumented detector holder}
The CaWO$_4$ sticks  (diameter 2.5\,mm, length $\sim$12\,mm) together with the scintillating and reflective foil surrounding the detectors provide a fully scintillating inner detector housing. This is crucial for the rejection of nuclear recoils from surface-alpha events which can mimic DM particle recoils. The corresponding MeV alpha particles  produce enough light in the scintillating  material to veto these dangerous backgrounds with high efficiency \cite{strauss:2014part1}.
 
Phonons from energy depositions in the CaWO$_4$ sticks to a certain extent  are transmitted via the point-like contact to the PD which results in a degraded signal in its TES. In particular, nuclear recoils  from surface-alpha events ($E=\mathcal{O}$(10-100\,keV)) occurring at the outward sides of the sticks might be detected as low-energy depositions of $\lesssim1$\,keV in the PD. For extremely low energy thresholds, as projected for CRESST-III, such events might limit the sensitivity. Furthermore, possible stress relaxation events related to the holding of the crystal might appear at lowest energies. 

To efficiently reject these event classes, the sticks are instrumented with TESs. Standard LD TESs (area of W-film: 0.30x0.08\,mm$^2$) are evaporated on Si carriers (3x3.5x0.4\,mm$^3$) which are glued onto the CaWO$_4$ sticks. Consequently, the full energy deposition is measured by the stick TES with a coincident (degraded) signal in the PD, and vice versa. The ratio of both signals can be used for a discrimination of any type of events occurring in the sticks. To reduce the number of readout channels, the 3 TES of the sticks holding the PD are connected in parallel to one SQUID\footnote{Every stick TES has an individual ohmic heater  for calibration and stabilization.}. 

\section{Results}
The CRESST-III prototype detector was operated at $\sim$10\,mK in a dilution refrigerator at MPI Munich (cryostat I, run127). In a test measurement with a 3-channel readout (PD, LD and sticks), an exposure of 69.5\,g-days was acquired. A $^{55}$Fe calibration source was placed inside the detector housing for an energy calibration of the PD. The observed overall event rate of about 5\,Hz allows a stable operation above ground, however affects the performance of the detectors (see below). In the following we will focus on the performance of the CaWO$_4$ PD, in particular its energy threshold, and the functionality of the novel instrumented detector holder.

\begin{figure}
\centering
\includegraphics[width=0.7\textwidth]{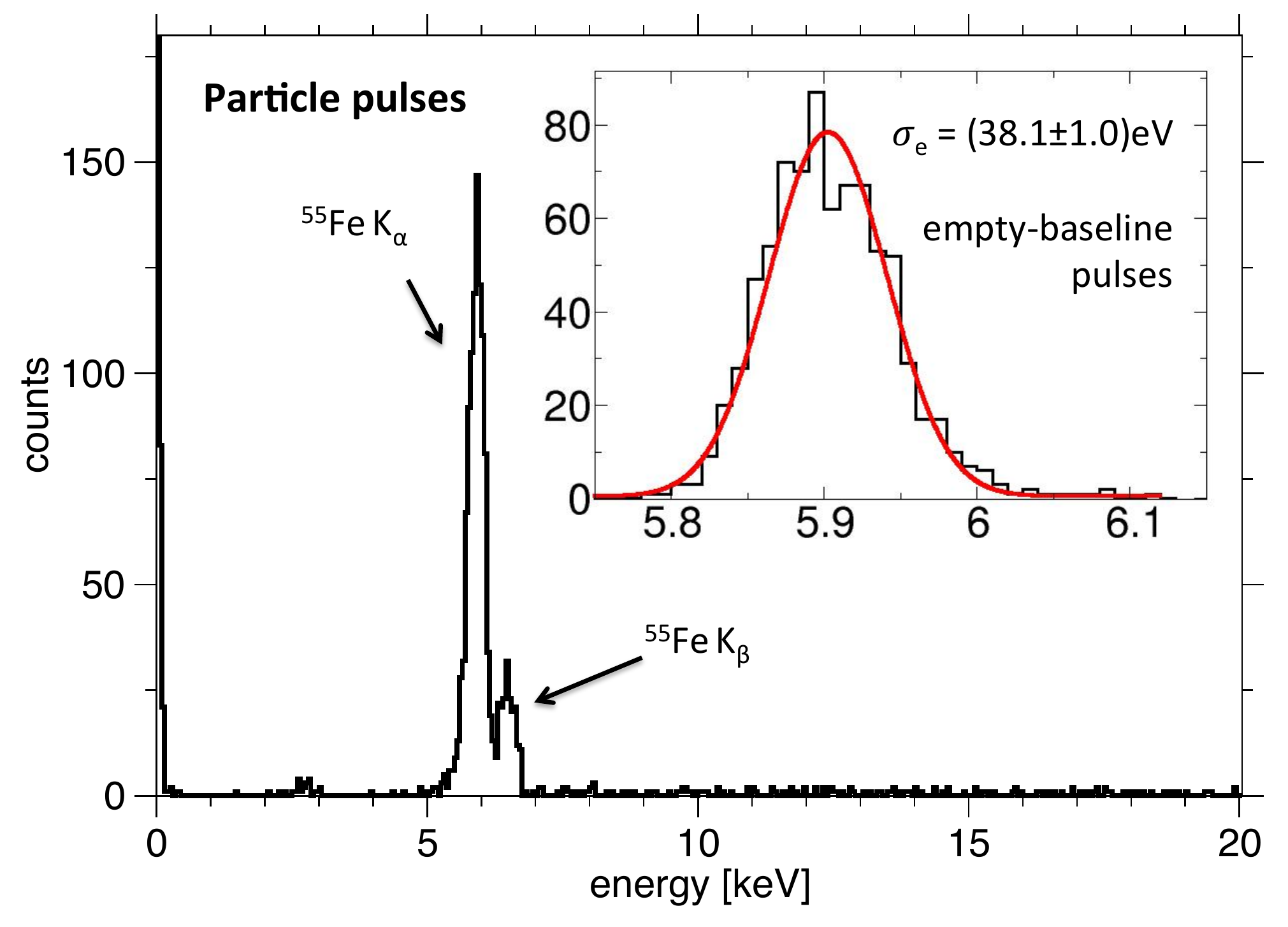}
\caption{Main frame: Calibration spectrum from a $^{55}$Fe source recorded with the CRESST-III prototype module. Due to a high pile-up rate in this measurement above ground, the resolution at $E\sim$6\,keV is slightly enhanced compared to $E=0$ (see text). Insert:  Energy distribution of artificial 5.90\,keV template pulses summed on recorded baseline samples (\textit{empty baselines}) determined by a template fit. The width of the distribution yields  a 5$\sigma$ energy threshold of $E_{th}=(190.6\pm5.2)\,$eV.}
\label{fig:calibration spectrum}
\end{figure}

The acquired energy spectrum is shown in Fig. \ref{fig:calibration spectrum} with the clearly identified $^{55}$Mn peaks at 5.9\,keV (K$_\alpha$) and 6.5\,keV (K$_\beta$) from the $^{55}$Fe source. The resolution at 5.9\,keV is $\sigma = (138.2\pm3.5)$\,eV which is expected to be enhanced compared to energies close to the threshold due to the high overall event rate\footnote{The high pile-up rate shifts the baseline of the pulses with time. Due to a non-linear detector response, a varying baseline negatively affects the resolution, increasingly at higher energies. For low event rates, only a minor energy dependence of the resolution is observed \cite{Angloher:2015ewa}.  }. 

\begin{figure}
\centering
\includegraphics[width=0.7\textwidth]{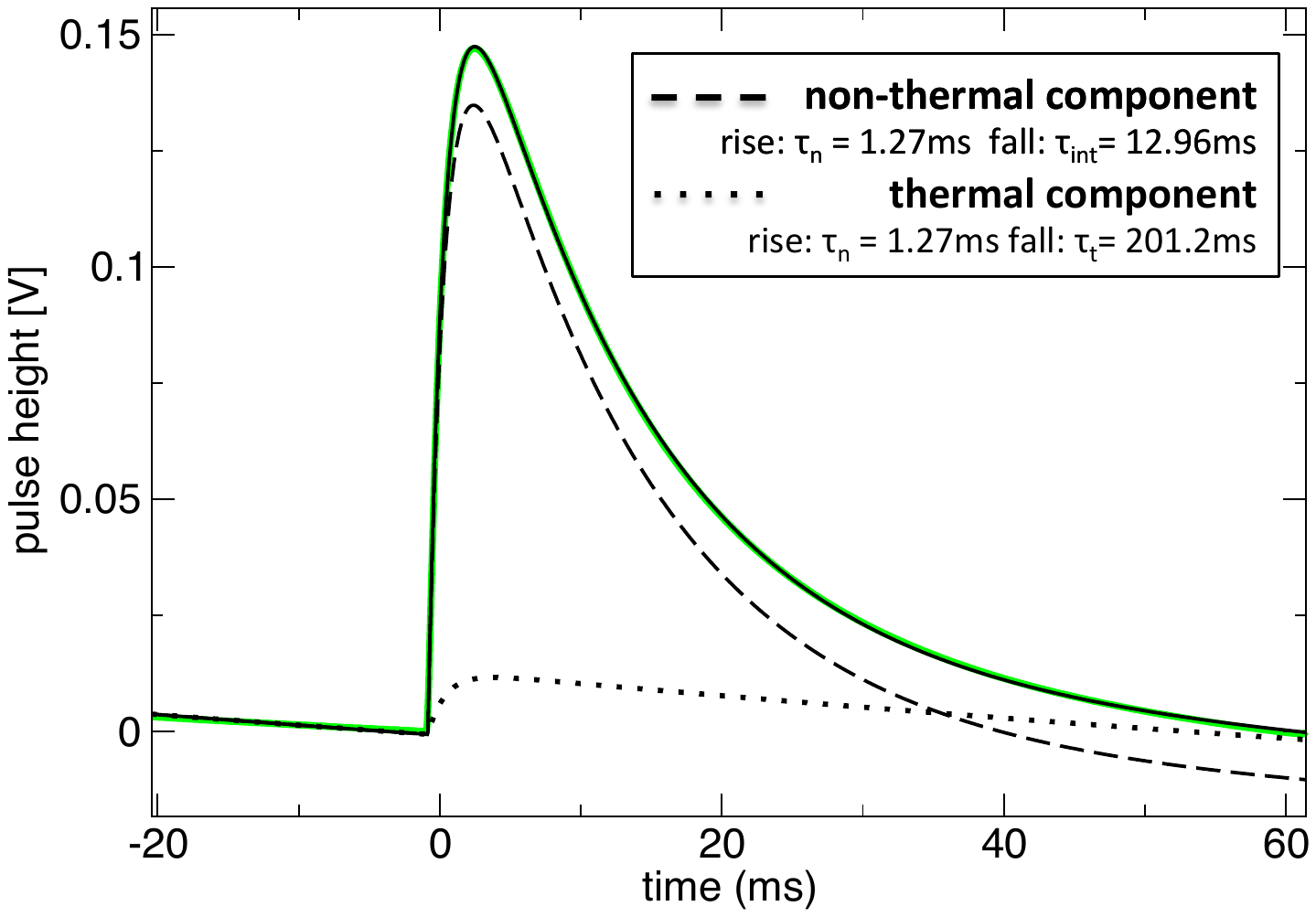}
\caption{Pulse template from 5.9\,keV  pulses from a $^{55}$Fe source acquired with the CRESST-III detector prototype (full green). It is fitted (full black) with the two-component pulse model (dashed and dotted black) developed in \cite{Probst:1995fk}. The resulting rise and decay times verify that the device is operated in the calorimetric mode (see text). The slightly tilted baseline	originates from the high event rate in the measurement. }
\label{fig:template}
\end{figure}

To derive the energy threshold of the prototype detector, the \textit{empty baseline method} is used which is the default method to validate the CRESST analysis \citep{Angloher:2014myn,Angloher:2015ewa}. Empty baselines, i.e. samples without any triggered pulses, are periodically acquired during data-taking. These samples are fitted with template pulses (see Fig. \ref{fig:template}) to derive the relevant noise level of the baseline. For technical reasons  the template pulse of a certain pulse height is summed to the empty baselines\footnote{The template fit optimizes pulse height and onset; for empty baselines the onset is not constrained and consequently the fit fails to reproduce the correct noise level. To fix the onset, a template of arbitrary (but fixed) pulse height is summed to the pulses. The choice of the pulse height does not influence the result of the baseline noise. }. The width of the reconstructed pulse height corresponds to the noise at $E=0$. Fig. \ref{fig:calibration spectrum} (inset) shows the results of the empty baseline method fitted by a Gaussian. A baseline noise of $\sigma=(9.79\pm0.26)\,$mV is derived which corresponds to $\sigma_e=(38.1\pm1.0)\,$eV and a 5$\sigma$ energy threshold of $E_{th}=(190.6\pm5.2)\,$eV. 

In the much more quiet CRESST setup, a baseline-noise level of typically $\sigma=$1.5-3.0\,mV is observed. For the CRESST-III prototype detector presented here, this corresponds to an energy threshold of $29-59$\,eV. 

The template pulse (from 5.9\,keV $^{55}$Fe pulses) is fitted by a two-component pulse model \cite{Probst:1995fk} to validate the calorimetric operation of the detector. The fit in Fig. \ref{fig:template} demonstrates that the template pulse can be well described by a dominant non-thermal phonon component which decays with the intrinsic time constant  of the TES, $\tau_{int}=12.96$\,ms and a thermal-phonon component which decays with the thermal decay time $\tau_t=201.2$\,ms  depending on the thermal coupling of the crystal to the heat bath. Both constituents rise with the lifetime $\tau_n=1.27\,$ms of the non-thermal phonons  in the detector. This is a clear proof that the detector works as a calorimeter (for details see \cite{Probst:1995fk}).

\begin{figure}
\centering
\includegraphics[width=.7\textwidth]{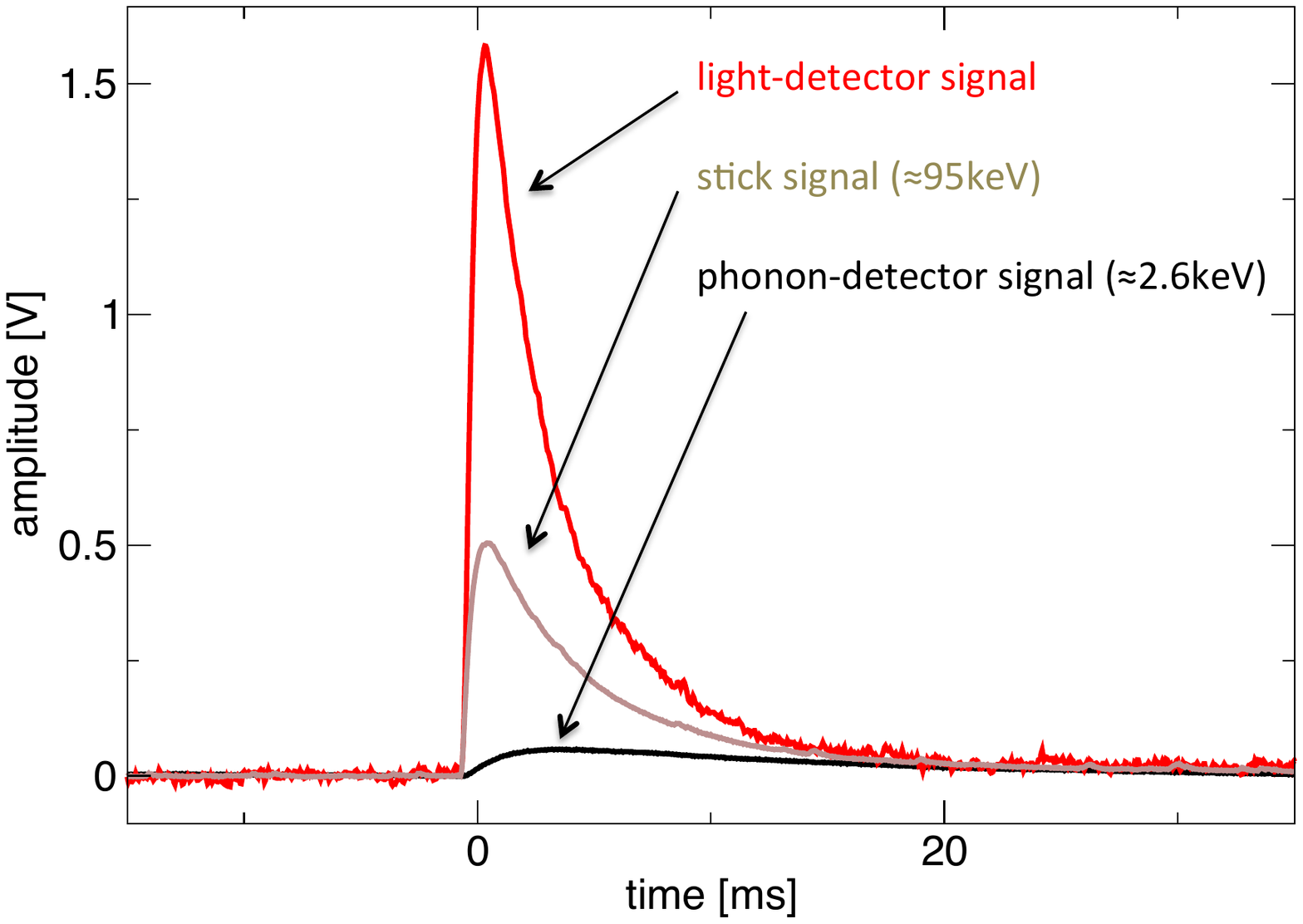}
\caption{Three-fold coincident pulses of a $\sim$95\,keV electron/gamma event occurring in one of the sticks holding the CaWO$_4$ crystal. In the PD, a degraded signal is observed which corresponds to a energy of $\sim$2.6\,keV. The scintillation light output of the stick is recorded with the LD.     }
\label{fig:iStick}
\end{figure}

The CRESST-III prototype module was operated for the first time in an instrumented detector holder. A detailed analysis of this system is beyond the scope of this paper, however, a few general aspects of the functionality are given here. All 3 CaWO$_4$ sticks which are equipped with TES could be run stably and an energy threshold of $\sim$1\,keV was achieved\footnote{A variation of about 20\% in the signal-to-noise ratio among the 3 sticks is observed.}. In Fig. \ref{fig:iStick}, a three-fold coincident pulse is shown, an energy deposition of $(95\pm10)\,$keV in one of the sticks probably originating from an electron/gamma event. Besides the signal in the stick TES (brown) and the corresponding scintillation light output of the CaWO$_4$ stick  measured by the LD (red), a degraded signal of $(2.6\pm0.1)$\,keV is detected in the PD TES. The ratio of the energy deposition in the stick and the degraded energy signal in the stick TES yields a degradation factor $D\sim$37 of the signal through the stick-crystal interface. The corresponding light signal originating from the electron/gamma event in the stick ($\sim$95\,keV) is accordingly higher by $D$ compared to a electron/gamma event of the same PD pulse height in the CaWO$_4$ crystal. Therefore, the light signal alone can be used for an efficient discrimination of  electron/gamma events occurring in the sticks. This event rejection was demonstrated in CRESST-II phase 2 \cite{strauss:2014part1}. For the rejection of nuclear-recoil backgrounds (e.g from surface-alpha reactions) and stress relaxations (no light signal) this method fails. 
However, using the instrumented detector holder such events can be rejected independently of the signal in the LD.  The lowest energy deposition in one of the sticks that produces a PD signal above threshold (190.6\,eV) is $\sim$7.1\,keV which is well above the threshold (${>}10\sigma$) of the stick-TESs. Hence, an efficient rejection of holder-related events down to the energy threshold of the PD is possible.

\section{Conclusion and Outlook}
A prototype CRESST-III module has successfully been  tested in a cryogenic setup above ground. Despite rather high cosmogenic and environmental backgrounds as well as a higher electronic noise level (by a factor of 3-6 compared to typical channels in the CRESST setup), a nuclear-recoil energy threshold of $E_{th}=(190.6\pm5.2)\,$eV was reached with the 24\,g CaWO$_4$ crystal. This is the lowest threshold reported for  direct DM search experiments. The result demonstrates, that the main design goal of CRESST-III phase 1, namely a threshold of $E_{th}=100$\,eV \cite{Angloher:2015eza}, should be easily achieved when the detectors are operated in the low-noise CRESST setup.
It was verified that the CRESST-III PD with its dedicated TES works in the calorimetric operation mode with a main pulse decay time of $\tau_{n}\sim$13\,ms. Furthermore, a novel instrumented detector holder was implemented and successfully operated. The CaWO$_4$ sticks are equipped with TESs which provide a  veto against the different types of events originating from the holding of the CaWO$_4$ target crystal. The results demonstrate an efficient rejection capability of these backgrounds down to the PD threshold. 

The detector prototype presented in this paper matches all requirements for low-mass DM search with CRESST-III phase 1. A dedicated study \cite{Angloher:2015eza} shows that with 10 such detectors operated for $\sim$1\,year, significant new parameter space in spin-independent DM particle-nucleus cross section can be probed and extended to masses of $\sim$0.1\,GeV/c$^2$. E.g., at a DM particle mass of 2\,GeV/c$^2$, a sensitivity improvement by $\sim$2 orders of magnitude to a cross section of $6\cdot10^{-6}$\,pb is expected.

At the time of writing, 10 CRESST-III detector modules were assembled and mounted in the CRESST setup   at the LNGS (Laboratori Nazionali del Gran Sasso) in Italy. Besides 7 CaWO$_4$ crystals of CRESST-II quality \cite{strauss:2014part2}, 3 CaWO$_4$ crystal of improved quality are installed. The latter are grown in the crystal growth facility at TUM from raw material which was purified by an extensive chemical treatment.  First data from CRESST-III phase 1 are expected in summer 2016. The presented detector design is also suited for the second phase of CRESST-III, in which $\sim$100  CaWO$_4$ crystals (24\,g each) of improved radiopurity \cite{Angloher:2015eza} are planned to be used. This future experiment will extend the sensitivity to low-mass DM particles by another 2 orders of magnitude and reach cross-sections close to that of coherent neutrino-nucleus scattering \cite{Guetlein2015}.  

\section*{Acknowledgments}
\begin{footnotesize}
This research was supported by the DFG cluster of excellence: “Origin and Structure of the Universe”, the DFG “Transregio 27: Neutrinos and Beyond”, the “Helmholtz Alliance for Astroparticle Phyiscs” and the “Maier-Leibnitz-Laboratorium” (Garching).
\end{footnotesize}

\section*{References}

\bibliography{VCI_proc}

\end{document}